\documentclass[aip, cha, twocolumn, groupedaddress, showpacs, floatfix, letterpaper, 10pt]{revtex4-1}

\usepackage{amsmath, amssymb, amsthm}
\usepackage{hyperref}
\usepackage[english]{babel}
\usepackage{graphicx}
\usepackage{subfigure}
\usepackage{times}

\newcommand{\T}{\mathbb{T}}
\newcommand{\R}{\mathbb{R}}
\newcommand{\Z}{\mathbb{Z}}

\newcommand{\C}{\mathbb{C}}

\newcommand{\vphi}{\theta}

\newcommand{\Ss}{\mathbf{S}}

\newcommand{\CIR}{\mathcal{C}}

\newcommand{\e}{\varepsilon}

\newcommand{\Splay}{\Theta^{\mathrm{splay}}}
\newcommand{\OmSplay}{\Omega^{\mathrm{splay}}}
\newcommand{\ldaSplay}{\lambda^{\mathrm{splay}}}
\newcommand{\Sync}{\Theta^{\mathrm{sync}}}
\newcommand{\OmSync}{\Omega^{\mathrm{sync}}}
\newcommand{\ldaSync}{\lambda^{\mathrm{sync}}}

\newcommand{\abs}[1]{\left|#1\right|}

\newcommand{\rset}[2]{\left\lbrace\, #1\,\left|\;#2\right.\right\rbrace}

\newcommand{\set}[2]{\rset{#1}{#2}}

\newcommand{\lmax}{\lambda_{\text{max}}}

\newcommand{\ud}{\textrm{d}}

\newtheorem{thm}{Theorem}

\theoremstyle{definition}

\theoremstyle{remark}

\begin{document}

\title{Chaos in generically coupled phase oscillator networks with nonpairwise interactions}

\author{Christian Bick, Peter Ashwin, and Ana Rodrigues}
\address{Centre for Systems, Dynamics and Control and Department of Mathematics, University of Exeter, Exeter EX4 4QF, UK}

\begin{abstract}
The Kuramoto--Sakaguchi system of coupled phase oscillators, where interaction between oscillators is determined by a single harmonic of phase differences of pairs of oscillators, has very simple emergent dynamics in the case of identical oscillators that are globally coupled: there is a variational structure that means the only attractors are full synchrony (in-phase) or splay phase (rotating wave/full asynchrony) oscillations and the bifurcation between these states is highly degenerate. Here we show that nonpairwise coupling---including three and four-way interactions of the oscillator phases---that appears generically at the next order in normal-form based calculations, can give rise to complex emergent dynamics in symmetric phase oscillator networks. In particular, we show that chaos can appear in the smallest possible dimension of four coupled phase oscillators for a range of parameter values.
\end{abstract}

\maketitle

\newcommand{\imagescaling}{0.8}

\begin{quotation}
In this paper, we show that symmetrically coupled phase oscillators with generic (but nonpairwise) interactions yield rich dynamics even for as few as $N=4$ oscillators. Although the lowest order approximation of a phase reduction of symmetrically coupled oscillators close to a Hopf bifurcation has only the Kuramoto--Sakaguchi first harmonic interaction terms, the next order includes generic terms with both second harmonic pairwise interactions and interactions of up to four phases~\cite{AR15}. The symmetries we consider imply the existence of invariant subspaces such that the ordering of phases is preserved~\cite{AS92}. In contrast to the Kuramoto--Sakaguchi equations, the additional nonpairwise interaction terms mean we can find attracting chaos for a range of normal form parameter values. As a consequence, the phase dynamics of generic weakly coupled oscillators will be quite rich and chaos can occur even for the phase dynamics in the weak coupling limit without amplitude degrees of freedom~\cite{NK94}.
\end{quotation}

\section{Introduction}

Recent advances in the understanding of the dynamics of coupled oscillators have shed light on the dynamical mechanisms involved in the emergence of collective behavior of oscillatory systems in nature and technology, including biology~\cite{Mirollo1990}, neuroscience~\cite{ACN15,Brown2003,Farmer02,Gray94}, chemistry~\cite{TTZS09,TNS12}, and physics~\cite{WS95,Acebron2005,Strogatz2000}. Even the dynamics of all-to-all coupled networks of identical oscillators can be extremely rich, including synchronization~\cite{Strogatz2000, Pikovsky2003}, clustering and slow switching~\cite{HMM93} and chaotic dynamics~\cite{HR92, NK94}. A general approach to understand clustering is to look at invariant synchrony subspaces~\cite{Aguiar2014} in coupled cell networks~\cite{Field04, Golubitsky2005, Golubitsky2006} and how these change as the network topology is varied~\cite{Aguiar2016}. Complicated dynamics arise already in very small networks~\cite{GucWor92, AP03} where bifurcations have been studied explicitly~\cite{Kuznetsov2016}.

If the coupling between~$N$ limit cycle oscillators is sufficiently weak then the dynamics can be approximated by a phase reduction. Based on the seminal work of Kuramoto~\cite{Kuramoto1975} to study the onset of synchronization in coupled oscillators~\cite{Strogatz2000}, phase oscillators whose phases $\theta_j\in \T=\R/2\pi\Z$ of oscillator~$j$ evolve according to
\begin{equation}
\dot\theta_j := \frac{\ud}{\ud t} \theta_j = \omega + \frac{1}{N} \sum_{k=1}^N g(\theta_k-\theta_j)=: F_j(\theta),
\label{eq:oscN}
\end{equation}
where~$\omega$ is the natural frequency of the oscillators and interaction is determined by the $2\pi$-periodic coupling (or phase interaction) function
\[g(\phi) = \sin(\phi+\alpha),\]
have been studied extensively~\cite{Acebron2005}. The dynamics given by~\eqref{eq:oscN}, the Kuramoto--Sakaguchi equations~\cite{SK86}, are degenerate: the dynamics are effectively two-dimensional and the only attractors are full synchrony or the splay phase oscillation depending on the value of~$\alpha$~\cite{Watanabe1994, Strogatz2000}. Thus, more complicated dynamics in~\eqref{eq:oscN} such as chaos is only possible for more general forms of coupling, for example by considering coupling functions~$g$ with higher harmonics. For $N=4$ phase oscillators---the smallest number where chaos can occur---no sign of chaos was found for coupling functions with two harmonics~\cite{ABB16} and the only known example of a coupling function giving rise to chaotic attractors has four nontrivial harmonics~\cite{Bick2011}. By contrast, two harmonics are sufficient to find chaos in networks~\eqref{eq:oscN} of $N=5$ phase oscillators~\cite{AOWT07}.

Recently, Ashwin and Rodrigues~\cite{AR15} showed that while a phase reduction of a generic fully symmetric system of oscillators close to a Hopf bifurcation to lowest order has the form~\eqref{eq:oscN}, the phase dynamics to the next order also contain higher order interaction terms that depend on three and four phases. More precisely, up to next order we have for coupling $0<\e\ll 1$ an invariant torus with phase dynamics given by
\begin{equation}\label{eq:principal1}
\begin{aligned}
\dot {\theta}_j &=  \tilde{\Omega}(\theta,\epsilon) 
+ \frac{\e}{N}\sum_{k=1}^N g_2(\theta_k-\theta_j) \\
&\qquad + \frac{\e}{N^2}\sum_{k,\ell=1}^{N}g_3(\theta_k+\theta_{\ell}-2\theta_j)\\
& \qquad+\frac{\e}{N^2}\sum_{k,\ell=1}^{N} g_4(2\theta_k-\theta_{\ell}-\theta_j) \\
&\qquad + \frac{\e}{N^3}\sum_{k,\ell,m=1}^{N} g_5(\theta_k+\theta_{\ell}-\theta_{m}-\theta_j)
\end{aligned}
\end{equation}
where $\tilde{\Omega}(\theta,\epsilon)$ is a symmetric function in the phases, the coupling function~$g_2$ has nontrivial first and second harmonics and the coupling functions $g_3, g_4, g_5$ that determine the nonpairwise interaction have a nontrivial first harmonic. Nonpairwise interaction in the phase model leads to novel nontrivial dynamical phenomena~\cite{AR15}; for example emergent quasiperiodicity of the mean field~\cite{RosPik2007, PikRos2009} which has been investigated explicitly~\cite{BP08}.

In this paper we study the dynamics of phase oscillator systems~\eqref{eq:principal1} that arise through the phase reduction of fully symmetric oscillators close to a Hopf bifurcation. In particular we discuss existence and stability of the in-phase (fully synchronous) and of the splay phase (rotating waves) solutions and calculate some bifurcations. The main result of this paper is to show that there are parameter values for coupling functions that give rise to positive maximal Lyapunov exponents for the dynamics of the phase differences of $N=4$ and $N=5$ oscillators, the former being the smallest number of oscillators for which chaotic dynamics in the truncated phase equations~\eqref{eq:principal1} can occur. In particular, we show evidence that chaos arises through period doubling and a Shilnikov scenario involving a saddle focus. Finally, we discuss the relationship of the chaos in the phase reduction and the dynamics of the full system of fully symmetric oscillators close to a Hopf bifurcation which can give rise to chaos even for three oscillators~\cite{AP03}.

This paper is organized as follows. In Section \ref{sec:Review} we review the results about the phase reduction of generic systems of oscillators with full symmetry close to a Hopf bifurcation~\cite{AR15}. In Section~\ref{sec:BasicsNonpairwise} we study basic properties of the resulting phase equations which we subsequently apply to small networks in Section~\ref{sec:SmallNets}. In Section~\ref{sec:Chaos} we give concrete examples of coupling functions that give rise to chaotic dynamics with positive maximal Lyapunov exponent before giving some concluding remarks.

\section{Nonpairwise Interaction in Phase Reduction near Hopf Bifurcations}
\label{sec:Review}

While the phase reduction of a system of weakly coupled oscillators close to a Hopf bifurcation yields phase dynamics with pairwise interaction terms to lowest order, it was recently shown in~\cite{AR15} that interaction terms of up to four phases can appear to next order if the coupling is very small compared to the distance from Hopf bifurcation. In this section we summarize these results and fix the notation.

Suppose we have~$N$ identical, symmetrically coupled dynamical systems with state~$x_k\in\R^d$ ($d\geq 2$) close to a Hopf bifurcation. Write $i=\sqrt{-1}$. Using equivariant bifurcation theory \cite{DR05} it is possible to write the system on a center manifold $(z_1,\ldots,z_N)\in\C^N$ where in the case $\lambda=\epsilon=0$ the center manifold in each coordinate $x_k$ is parametrized by $z_k$. This system on the center manifold is
\begin{equation*}
\begin{split}
\dot{z}_1 &= f_{\lambda}(z_1)+ \epsilon g_{\lambda}(z_1;z_2,\ldots,z_N)+O(\epsilon^2)\nonumber\\
&\ \,\vdots \label{eq:CNsystem}\\
\dot{z}_N &= f_{\lambda}(z_N)+ \epsilon g_{\lambda}(z_N;z_1,\ldots,z_{N-1})+O(\epsilon^2)\nonumber
\end{split}
\end{equation*}
where $z \in {\mathbb C}^{N}$ and we note the right hand sides can be chosen to be of smoothness~$C^r$, with~$r$ arbitrarily large, in a neighborhood of the bifurcation. The conditions for Hopf bifurcation mean that for \eqref{eq:CNsystem} we have $f_{0}(0)=0$ and the derivative $\ud f_{0}(0)$ has a pair of purely imaginary eigenvalues~$\pm i\omega$ that pass transversely through the imaginary axis with non-zero speed on changing $\lambda$. Let $\Ss_N$ denote the group of permutations of~$N$ symbols which acts on~$\C^N$ by permutation of coordinates, that is, if $\sigma \in \Ss_N$ then
\begin{equation}\label{eqaction1}
\sigma (z_1,\ldots, z_N) = (z_{\sigma^{-1}(1)}, \ldots, z_{\sigma^{-1}(N)}),
\end{equation}
where $(z_1,\ldots,z_N) \in \C^N$. So $g_{\lambda}(z_1;z_2,\ldots,z_N)$ is symmetric under all permutations of the last~$N-1$ arguments that fix the first.

As shown in~\cite{AR15}, taking higher orders into account we obtain an invariant torus with phase dynamics~\eqref{eq:principal1}. More precisely, with
\begin{align}
F^{(2)}_j(\theta) &= \frac{1}{N}\sum_{k=1}^N g_2(\theta_k-\theta_j),\\
\begin{split}
F^{(3)}_j(\theta) &= \frac{1}{N^2}\sum_{k,\ell=1}^{N}g_3(\theta_k+\theta_{\ell}-2\theta_j)\\
&\qquad + \frac{1}{N^2}\sum_{k,\ell=1}^{N} g_4(2\theta_k-\theta_{\ell}-\theta_j),
\end{split}\\
F^{(4)}_j(\theta) &= \frac{1}{N^3}\sum_{k,\ell,m=1}^{N} g_5(\theta_k+\theta_{\ell}-\theta_{m}-\theta_j).
\end{align}
the phase dynamics are
\begin{equation}\label{eq:principal}
\begin{aligned}
\dot {\theta}_j &= \tilde{\Omega}(\theta,\epsilon) + \e\left(F^{(2)}_j(\theta)+F^{(3)}_j(\theta)+F^{(4)}_j(\theta)\right)
\end{aligned}
\end{equation}
where $\tilde{\Omega}(\theta,\epsilon)$ is a symmetric function in the phases and the interactions between the phases is given by
\begin{equation}
\begin{split}
g_2(\phi)&= \xi_1^0 \cos (\phi+\chi_1^0)+\lambda \xi_1^1 \cos (\phi+\chi_1^1)\\
&\qquad+ \lambda\xi_2^1\cos (2\phi+\chi_2^1)\\
g_3(\phi)&=  \lambda\xi_3^1 \cos (\phi+\chi_3^1)\\
g_4(\phi)&= \lambda\xi_4^1 \cos (\phi+\chi_4^1)\\
g_5(\phi)&=  \lambda\xi_5^1 \cos (\phi+\chi_5^1)
\end{split}
\label{eq:coupform2}
\end{equation}
for some constant coefficients $\xi_i^j$ and $\chi_i^j$. Note that \eqref{eq:oscN} corresponds to the special case of \eqref{eq:principal} where $F_j^{(k)}\equiv 0$ for $k=3$ and $k=4$. More precisely, \cite{AR15} proves that \eqref{eq:principal} is the next approximate truncation of the normal form after Kuramoto--Sakaguchi, in the following sense.

\begin{thm}{\bf \cite[Theorem~3.2]{AR15}}\label{thm:main}
	Consider system \eqref{eq:CNsystem} with $\Ss_N$-symmetry (for fixed~$N$) such that the $N$ uncoupled systems ($\epsilon=0$) undergo a generic supercritical Hopf bifurcation on $\lambda$ passing through $0$. There exists $\lambda_0>0$ and $\epsilon_0=\epsilon_0(\lambda)$ such that for any $\lambda \in (0,\lambda_0)$ and $|\epsilon|<\epsilon_0(\lambda)$ the system \eqref{eq:CNsystem} has an attracting $C^r$-smooth invariant $N$-dimensional torus for arbitrarily large~$r$. Moreover, on this invariant torus, the phases $\theta_j$ of the flow can be expressed as \eqref{eq:principal}
for fixed $0<\lambda<\lambda_0$ in the limit $\epsilon\rightarrow 0$, where $\tilde{\Omega}(\phi,\epsilon)$ is independent of $j$ and $g_{2,\ldots,5}$ are given by \eqref{eq:coupform2}.
	The constants $\xi_i^j$ and $\chi_i^j$ are generically non-zero. 
	The error term satisfies
	$\tilde{g}= O(\lambda^2)$
	uniformly in the phases $\phi_k$. The truncation to \eqref{eq:principal} on removing the error terms $\tilde{g}$ and $O(\epsilon^2)$ terms is valid over time intervals $0<t<\tilde{t}$ where $\tilde{t}=O(\epsilon^{-1}\lambda^{-2})$ in the  limit $0< \epsilon\ll\lambda\ll 1$. 
\end{thm}

As noted in~\cite{AR15}, the presence of nonpairwise coupling can give rise to new phenomena in terms of bifurcation of two-cluster states. In the following we explore the effect of nonpairwise coupling further, in particular with respect to small networks and show that chaotic attractors may appear at this order of truncation.

\section{Dynamics of Phase Oscillators with Nonpairwise Coupling}
\label{sec:BasicsNonpairwise}

We discuss in this section the behavior of the system of phase oscillators \eqref{eq:principal}---to some extent this is similar to the case of pairwise coupling~\eqref{eq:oscN} in that there is a strong structure of invariant subspaces imposed by the permutation symmetries.

\subsection{Symmetric phase oscillator dynamics}

\begin{figure}%
	\centerline{
		\includegraphics[width=\columnwidth]{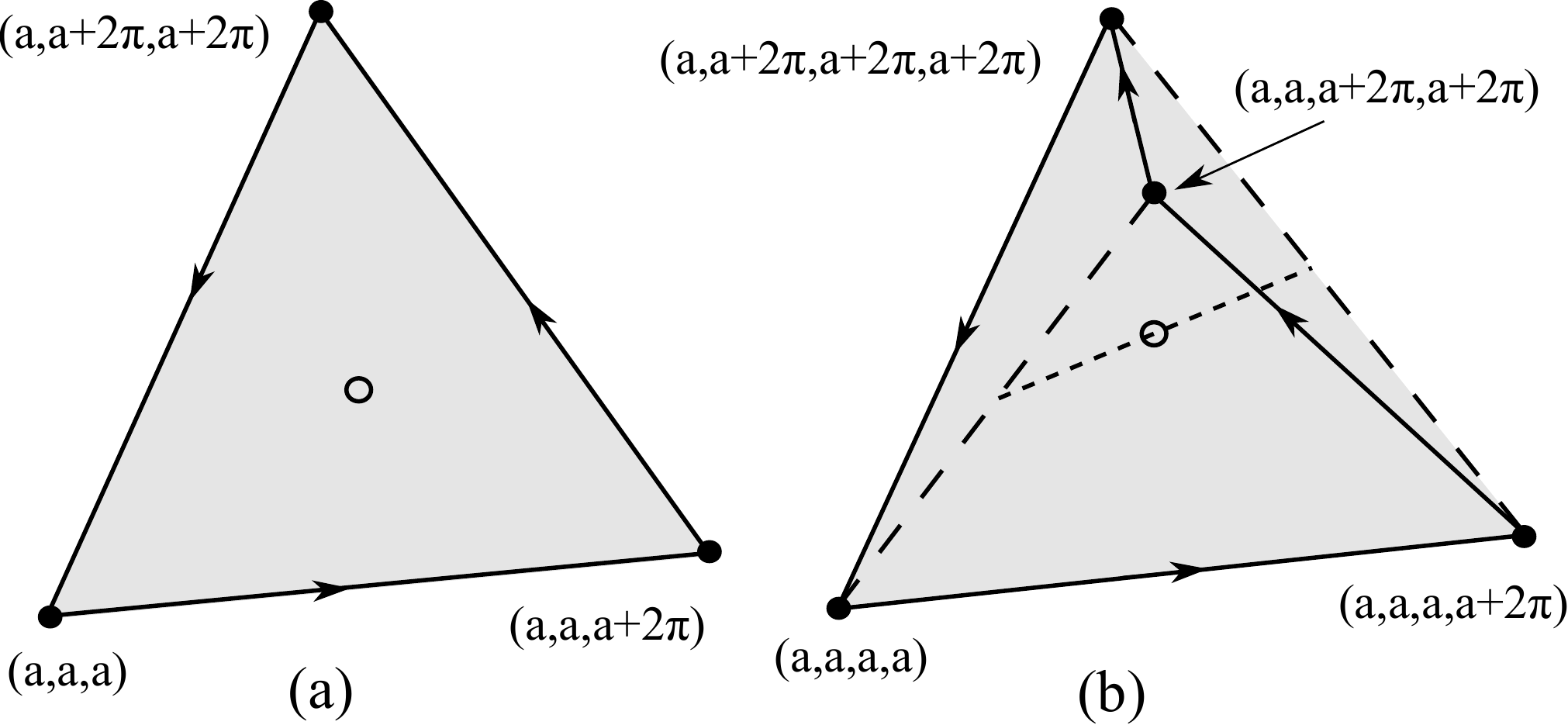}
	}%
	\caption{\label{fig:CIR}(Adapted from Ref.~\cite{ABB16}) Structure of the canonical invariant region $\CIR$ for $N=3$ and $N=4$ (see \cite{AS92}). Panels~(a,b) show~$\CIR$ as an orthogonal projection of into~$\R^2$ and~$\R^3$, respectively. The edges of~$\CIR$ for (a) and the faces of~$\CIR$ for (b) are points with~$\Ss_2$ isotropy. The filled circles represent different points on the lift that that correspond to fully synchronous oscillation; the open circle represents the splay phase oscillation in~$\CIR$. In (a) the solid lines have isotropy $\Ss_3\times\Ss_1$ while the long dashes have isotropy $\Ss_2\times\Ss_2$. The short dashed lines have isotropy $(\Ss_1)^2\times_S \Z_2$---typical points being $(a,b,a+\pi,b+\pi)$. In each case there is a residual $\Z_{N-1}$ symmetry indicated by the arrows in~(b) and $(N-1)!$ symmetric copies of~$\CIR$ pack a generating region for the torus.}%
\end{figure}

By considering a projection $\Pi:\T^N\rightarrow \T^{N-1}$ that maps the $\T$-orbits onto points, the generalized system \eqref{eq:principal} (we assume $\epsilon=1$ from hereon) reduces to phase differences on $\T^{N-1}$. The fixed point subspaces where two of the phases are identical forms a partition of~$\T^{N-1}$ into connected components that are all symmetric images of the \emph{canonical invariant region} (or CIR)~\cite{AS92} given by
\begin{equation}
	\CIR = \set{\theta\in\T^N}{0=\theta_1<\theta_2<\dotsb<\theta_N<2\pi}.
\end{equation}
Note that this region is invariant for any phase oscillator system with full permutation symmetry; moreover the region has symmetry $\Z_N=\Z/N\Z$ generated by
\begin{equation}\label{eq:RemSym}
	\tau: (0, \theta_2, \dotsc, \theta_N)\mapsto(0, \theta_3-\theta_2, \dotsc, \theta_N-\theta_2, 2\pi-\theta_2).
\end{equation}
For $N=3$ and $N=4$ the CIR is illustrated in Figure~\ref{fig:CIR}---the boundaries of $\CIR$ are invariant for dynamics.

For any partition of $N=m_1+\cdots+m_\ell$, $\ell\geq 2$, the CIR~$\CIR$ can be partitioned into \emph{$\ell$-cluster states} with isotropy $\Ss_{m_1}\times\cdots\times\Ss_{m_\ell}$ with~$\ell$ clusters of~$m_k$ oscillators at the same phase. More generally, there are invariant subspaces (rotating blocks) that have a phase shift symmetry as well as clustering; see \cite{AS92}. The paper \cite{ABB16} explores the dynamics of pairwise coupling~\eqref{eq:oscN} for general second harmonic coupling in the cases $N=3$ and $N=4$ and in particular, no evidence is found of chaotic attractors for either case.

This structure is instructive in that is places limits on where any chaotic behavior can be found. As pointed out in \cite{Bick2011}, the fact that $N=3$ reduces to planar dynamics on $\CIR$ immediately implies there cannot be any chaotic behavior in this case, while for $N=4$, if there is any chaotic behavior it must include points that have trivial isotropy (i.e., that are not in any invariant subspace, as these all have dimension two or less).

\subsection{Fully synchronized and splay phase: existence and stability}

We recall from \cite{AS92,ABB16} two important periodic solutions that are guaranteed to exist for \eqref{eq:principal}. The in-phase (fully phase synchronized) oscillation corresponds to the solution $\Sync = (\OmSync t, \dotsc, \OmSync t)$ for some $\OmSync$ while the splay phase (rotating wave, $\Z_N$ symmetric solution) corresponds to the solution $\Splay$ with $\Splay_j=\OmSplay t + \frac{2\pi}{N}(j-1)$.

The stability for in-phase oscillation can be computed from the Jacobian of \eqref{eq:principal}, namely
\begin{equation}
\label{eq:jacobian}
\begin{split}
J_{jk}(\vphi)&=\frac{\partial}{\partial \vphi_k}F_j(\vphi)\\
&=\frac{\partial }{\partial \vphi_k}\left(F^{(2)}_j(\vphi)+F^{(3)}_j(\vphi)+F^{(4)}_j(\vphi)\right).
\end{split}
\end{equation}
{\allowdisplaybreaks
For $k\neq j$ we have
\begin{align*}
\frac{\partial}{\partial\vphi_k} F^{(2)}_j &= \frac{1}{N}g_2'(\vphi_k-\vphi_j),\\
\begin{split}
\frac{\partial}{\partial\vphi_k} F^{(3)}_j  &= \frac{1}{N^2}\sum_{\ell=1}^{N}\Big(2 g_3'(\vphi_k+\vphi_\ell-2\vphi_j)-g_4'(2\vphi_\ell-\vphi_k-\vphi_j)\\
&\hspace{0.3\linewidth}+2g_4'(2\vphi_k-\vphi_\ell-\vphi_j)\Big),
\end{split}\\
\begin{split}
\frac{\partial}{\partial\vphi_k} F^{(4)}_j  &= \frac{1}{N^3}\sum_{\ell,m=1}^{N}\Big(2g_5'(\vphi_j+\vphi_k-\vphi_\ell-\vphi_m)\\
&\hspace{0.3\linewidth}-g_5'(\vphi_j+\vphi_\ell-\vphi_k-\vphi_m)\Big),
\end{split}
\end{align*}
}
where $g_k'$ denotes the derivative of $g_k:\R\to\R$ with respect to its argument.
Because of the phase shift symmetry, the Jacobian of the vector field $F$ has a zero eigenvalue with eigenvector  $(1, \dotsc, 1)$. This implies that
\[
\frac{\partial}{\partial\vphi_j} F_j(\vphi) = -\sum_{k\neq j}\frac{\partial}{\partial\vphi_k} F_j(\vphi).
\]

In the case of in-phase oscillation for $j\neq k$ the coefficients of the Jacobian simplify to
\begin{equation}
J_{jk} = \frac{1}{N}\left(g_2'(0)+ 2g'_3(0)+g'_4(0)+g'_5(0)\right)
\end{equation}
and $J_{kk}= - \sum_{q\neq k}J_{kq}$. Hence for this case there is a zero eigenvalue and $N-1$ other eigenvalues that are given by
\begin{equation}
\label{eq:inphaseeig}
\ldaSync=-g_2'(0)- 2g'_3(0)-g'_4(0)-g'_5(0).
\end{equation}
Hence the in-phase oscillation will lose stability when~$\ldaSync$ passes from negative to positive in a highly degenerate symmetric bifurcation (see \cite{AS92} for more discussion of some of the branches and global attractors that may generically appear at such a bifurcation). This can be expressed as a weighted sum of the derivatives of $g_{k}(0)$.

In the case of the splay phase oscillation we can compute, for $j\neq k$, that
\begin{align*}
J_{jk} &= \frac{1}{N} g_2'((k-j)\omega)\\
&\qquad+\frac{1}{N^2} \sum_{\ell=1}^N \Big(2g_3'((k+\ell-2j)\omega)-g_4'((2\ell-k-j)\omega)\\
&\hspace{0.4\linewidth}+2g_4'((2k-\ell-j)\omega)\Big)\\
&\qquad+\frac{1}{N^3} \sum_{\ell,m=1}^N \Big(2g_5'((j+k-\ell-m)\omega)\\
&\hspace{0.4\linewidth}-g_5'((j+\ell-k-m)\omega)\Big).
\end{align*}
Note that although this is not clear from this expression, there should be circulant structure $J_{j+\ell\,k+\ell}=J_{jk}$ for this matrix which implies that the non-zero eigenvalues will generically be complex except (in the case of $N$ even) for a single real eigenvalue. We do not compute these eigenvalues in their full generality but give them for the special cases $N=2,3,4$ in the next section.


\section{Dynamics of Small Networks with Nonpairwise Coupling}
\label{sec:SmallNets}
Since the nonpairwise coupling involves combinations of three and four phases, the dynamics for $N=2$ and $N=3$ oscillators reduce to coupling of simpler form. In particular, for $N=2$ oscillators~\eqref{eq:principal} reduces to pairwise coupling~\eqref{eq:oscN} with coupling function $g(\vphi)=\xi_1 \cos(\vphi+\chi_1)+\xi_2\cos(2\vphi+\chi_2)$. Similarly, for $N=3$ we can express the contributions of~$F_j^{(4)}$ to the dynamics in terms of just pairwise and three-phase interactions. Consequently, $N=4$ is the simplest case where all higher order interaction terms in~\eqref{eq:principal} are nontrivial.

In the following, we consider coupling functions of the form of \eqref{eq:coupform2} where $\lambda$ is fixed, i.e.,
\begin{equation}
\begin{split}
g_2(\phi)&= \xi_1 \cos(\phi+\chi_1)+\xi_2\cos(2\phi+\chi_2)\\
g_3(\phi)&= \xi_3 \cos(\phi+\chi_3)\\
g_4(\phi)&= \xi_4 \cos(\phi+\chi_4)\\
g_5(\phi)&= \xi_5 \cos(\phi+\chi_5)
\end{split}
\label{eq:coupform3}
\end{equation}
where for general $N$ the function~$g_2$ determines pairwise, $g_3, g_4$ triplet and $g_5$ quadruplet interaction. The cases $N=2$ and $3$ are special case as we now discuss. We note that the eigenvalue~\eqref{eq:inphaseeig} that determines the stability of the in-phase solution~$\Sync$ evaluates to
\begin{equation}
\label{eq:synceigs}
\begin{split}
\ldaSync &= \xi_1\sin(\chi_1)+2\xi_2\sin(\chi_2)\\
&\qquad +2\xi_3\sin(\chi_3)+\xi_4\sin(\chi_4)+\xi_5\sin(\chi_5).
\end{split}
\end{equation}

\subsection{Dynamics of $N=2$ or $3$ oscillators}
For $N=2$ note that \eqref{eq:principal} with coupling \eqref{eq:coupform3} can be written as \eqref{eq:oscN} where
\begin{align*}
g(\phi)&:=g_2(\phi)+g_3(\phi)+\frac{1}{2}g_3(2\phi)\\
&\qquad+\frac{1}{2}g_4(2\phi)+\frac{1}{4}\left(g_4(\phi)+g_4(-\phi)\right)\\
&\qquad+\frac{3}{4}g_5(\phi)+\frac{1}{4}\left(g_5(2\phi)+g_5(-\phi)\right).
\end{align*}
Combining these terms means that $g(\phi)$ is of the second harmonic form as studied in \cite{ABB16}. The bifurcation of in-phase oscillations is where \eqref{eq:synceigs} gives zero eigenvalue, i.e., where
\begin{equation}
\label{eq:inphaseN234}
\begin{split}
&\xi_1 \sin (\chi_1)+2 \xi_2 \sin(2\chi_2)+2\xi_3\sin(\chi_3)\\
&\hspace{0.3\linewidth}+\xi_4\sin(\chi_4)+\xi_5\sin(\chi_5)=0
\end{split}
\end{equation}
while the splay phase for $N=2$ corresponds to the antiphase state: this bifurcates where
\begin{equation}
\label{eq:splayphaseN2}
-\xi_1 \sin(\chi_1)+2 \xi_2 \sin(2\chi_2)+\xi_4\sin(\chi_4)+\xi_5\sin(\chi_5)=0.
\end{equation}

In the case of $N=3$ one can in principle subsume the terms $g_5$ into the terms $g_{2,3,4}$. The bifurcation of in-phase oscillations similarly occurs where \eqref{eq:inphaseN234}. The splay phase has eigenvalues that can be computed as
\begin{align*}
\ldaSplay&=-\xi_2 \sin(\chi_2) - \frac{1}{2}\xi_1\sin(\chi_1)\\
&\qquad \pm \frac{i}{2}\abs{2\xi_2\cos(\chi_2)-\xi_1 \cos(\chi_1)}
\end{align*}
which we note only depends on~$g_2$. Moreover these eigenvalues are complex unless $\xi_1\cos(\chi_1)=2\xi_2\cos(\chi_2)$ and there is a Hopf bifurcation of the splay phase for $N=3$ when
\begin{equation}
\label{eq:splayphaseN3}
\xi_2 \sin(\chi_2) + \frac{1}{2}\xi_1\sin(\chi_1)=0.
\end{equation}

\subsection{Dynamics of $N=4$ oscillators} 
Turning to $N=4$, there is similarly a bifurcation of in-phase solutions where \eqref{eq:inphaseN234} is satisfied. The splay phase for $N=4$ has eigenvalues
\[
\ldaSplay\in\left\{-2\xi_2 \sin(\chi_2), \frac{\xi_1}{2}\left(-\sin(\chi_1)+i\left|\cos(\chi_1)\right|\right) \right\}
\]
meaning there is a steady bifurcation of splay phase when
\begin{equation}
\label{eq:splaysteadyN4}
\xi_2\sin(\chi_2)=0
\end{equation}
while there is a Hopf bifurcation of splay phase (as long as $\cos(\chi_1)\neq 0$) when
\begin{equation}
\label{eq:splayHopfN4}
\xi_1\sin(\chi_1)=0.
\end{equation}
Moreover, for $N=4$ the dynamics on the one-dimensional invariant subspace with isotropy $\Z_2$---the  points $(0,\theta,\pi,\theta+\pi)$---is given by
\[
\dot{\theta} = \xi_2 \sin(\chi_2)\sin (2\theta)
\]
with bifurcation of splay phase $\theta=\frac{\pi}{2}$ at~\eqref{eq:splaysteadyN4}. For points $(0,0,\theta,\theta)$ with isotropy $(\Ss_2)^2$ we have
\begin{align*}
\dot{\theta}&= c_1\sin(\theta)+c_2\sin(2\theta)
\end{align*}
with
\begin{align*}
c_1 &= \xi_1\sin(\chi_1)+\xi_3\sin(\chi_3)+\frac{1}{2}\xi_5\sin(\chi_5),\\
\begin{split}
c_2 &= \xi_2\sin(\chi_2)+\frac{1}{2}\xi_3\sin(\chi_3)\\
&\qquad+\frac{1}{2}\xi_4\sin(\chi_4)+\frac{1}{4}\xi_5\sin(\chi_5).
\end{split}
\end{align*}
In this invariant subspace there is a bifurcation of the in-phase oscillation ($\theta=0$), as expected, at \eqref{eq:inphaseN234} while there is a bifurcation of the antiphase state $(0,0,\pi,\pi)$ at
\begin{equation}
\label{eq:antiphaseN4}
\xi_1\sin(\chi_1)-2\xi_2\sin(\chi_2)-\xi_4\sin(\chi_4)=0.
\end{equation}


\section{Chaos in Small Networks with Nonpairwise Coupling}
\label{sec:Chaos}	

Since the reduced system of~$N$ oscillators evolves on~$\T^{N-1}$, only networks of $N\geq 4$ oscillators can exhibit chaotic dynamics. We calculate the expansion of a perturbation along a trajectory by integrating the variational equations
\[\dot v_j = \sum_{k=1}^N J_{jk}(\theta(t))v_k\]
numerically along a solution $\theta(t)$ of \eqref{eq:principal} with Jacobian \eqref{eq:jacobian}: see for example~\cite{Bick2011} for more details. For a generic choice of~$v_k(0)$, we expect $v_k(t)$ to grow exponentially fast at the rate of the maximal Lyapunov exponent~$\lmax$, and for typical choices of initial conditions in the basin of the attractor, this growth rate will be independent of initial condition; by integrating the variational equations for the phase difference only we do not calculate the trivial zero Lyapunov exponent.

\begin{figure*}
\subfigure[\ Chaos in~$\CIR$ for phase shift parameters $\chi = (0.1, 0.267, 0, 1.5, 0)$]{\raisebox{0.15cm}{\includegraphics[scale=\imagescaling]{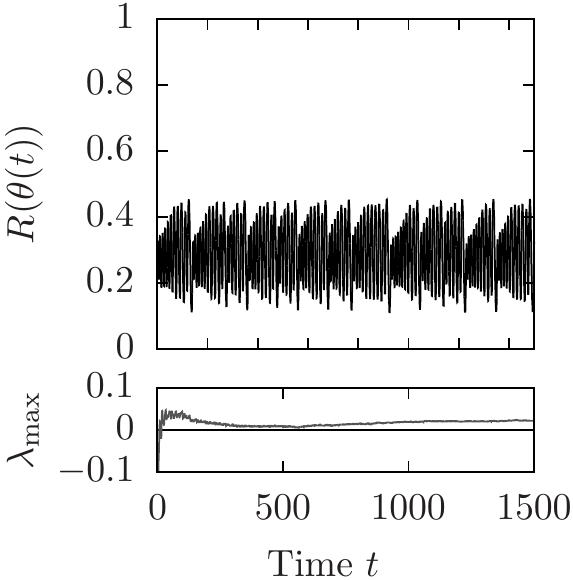}}\quad\raisebox{1cm}{\includegraphics[scale=0.65]{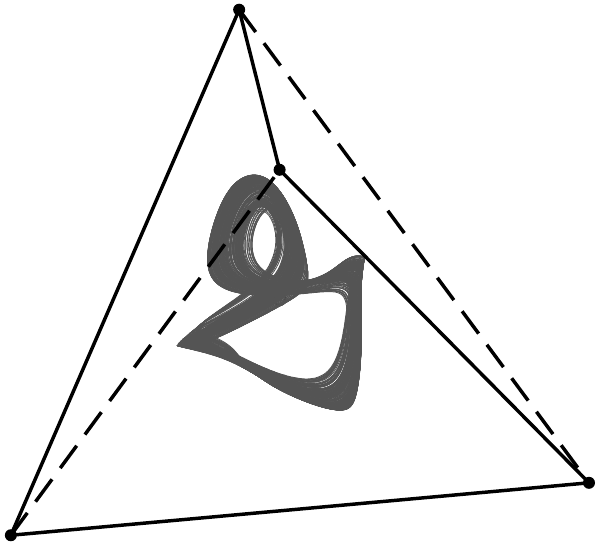}}}\qquad
\subfigure[\ $\lmax$ for varying parameters $\chi= (\chi_1, \chi_2, 0, 1.5, 0)$]{\includegraphics[scale=\imagescaling]{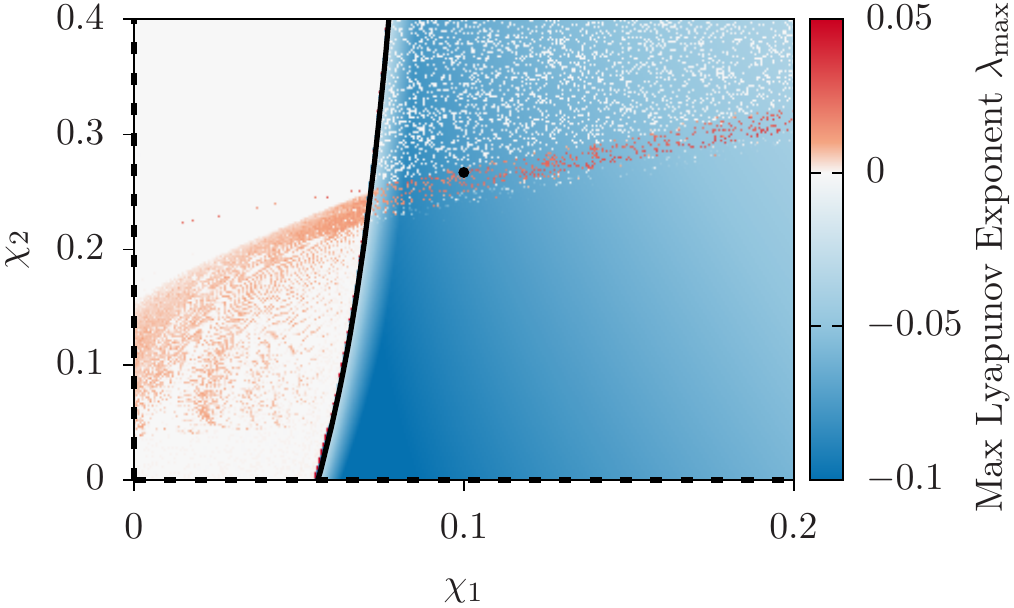}}
\caption{\label{fig:ChaosN4}
Chaotic attractors exist in~$\CIR$ for networks of $N=4$ oscillators~\eqref{eq:principal} with coupling function~\eqref{eq:coupform3}. Panel~(a) shows the dynamics for $\chi = (0.1, 0.267, 0, 1.5, 0)$ with the chaotic fluctuations of the absolute value of the order parameter~\eqref{eq:OrdParam} and the convergence of~$\lmax$ on the left and the attractor in terms of phase differences $\psi_k = \theta_k-\theta_1$ in~$\CIR$ on the right. The line styles for the cluster states on the boundary of~$\CIR$ are as in Figure~\ref{fig:CIR}. Panel (b) shows a region in parameter space where trajectories give rise to positive maximal Lyapunov exponents as parameters $\chi= (\chi_1, \chi_2, 0, 1.5, 0)$ are varied. The coloring indicates the maximal Lyapunov exponent which is negative if the trajectory it converges to a stable equilibrium (on $\T^{N-1}$), zero if it converges to a limit cycle, and positive if trajectories separate exponentially. As initial conditions are chosen randomly, speckled regions indicate bistability. Black lines indicate bifurcation of equilibria: the splay phase (dashed) and an equilibrium on the boundary of~$\CIR$ (solid). A black dot indicates the choice of parameters in Panel~(a).
\bigskip}
\end{figure*}

\subsection{Chaos in networks of $N=4$ oscillators}
For appropriately chosen parameters, networks of generically coupled phase oscillators~\eqref{eq:principal} with coupling functions~\eqref{eq:coupform3} give rise to positive maximal Lyapunov exponents. With fixed Fourier coefficients
\begin{align}
\xi &= (-0.3, 0.3, 0.02, 0.8, 0.02)
\end{align}
we explore the dynamics depending on the phase shifts~$\chi$. The absolute value of the order parameter,
\begin{equation}\label{eq:OrdParam}
R(\theta) = \abs{\frac{1}{N}\sum_{k=1}^{N}\exp(i\theta_k)},
\end{equation}
gives information about the synchronization of the oscillators, that is, $R(\Sync(t))=1$ and $R(\Splay(t))=0$. Figure~\ref{fig:ChaosN4}(a) shows chaotic dynamics for phase shifts $\chi = (0.108, 0.27, 0, 1.5, 0)$ within~$\CIR$ that give rise to positive maximal Lyapunov exponents. Integrating the system~\eqref{eq:principal}\footnote{Single trajectories were calculated in MATLAB using the standard adaptive Runge--Kutta scheme with relative and absolute error tolerances chosen conservatively at~$10^{-9}$ and~$10^{-11}$ respectively. For the parameter scans, the maximal Lyapunov exponents were calculated by integrating the system for $T=10^5$ time units using a standard fourth order Runge--Kutta scheme with fixed time step of $\Delta t = 10^{-2}$.} for varying parameters $\chi_1, \chi_2$ reveals a region 
with 
random initial condition reveals a region in parameter space where trajectories give positive maximal Lyapunov exponents; see Figure~\ref{fig:ChaosN4}(b). This region relates to the bifurcation lines $\chi_1=0$, given by~\eqref{eq:splaysteadyN4}, and $\chi_2=0$, given by~\eqref{eq:splayHopfN4}, of the splay phase. More specifically, numerical continuation of the branch of periodic solutions which arises in the Hopf bifurcation of~$\Splay$ in AUTO for fixed $\chi_1=0.1$ and decreasing $\chi_2$ from $0.3$ towards the parameter values of Figure~\ref{fig:ChaosN4} shows subsequent period doubling bifurcations (not shown). The bifurcation of a (relative) equilibrium on the boundary of~$\CIR$ induces bistability with the attractors in the interior of~$\CIR$.

\begin{figure}
\subfigure[\ Chaotic attractor]{\includegraphics[scale=0.65]{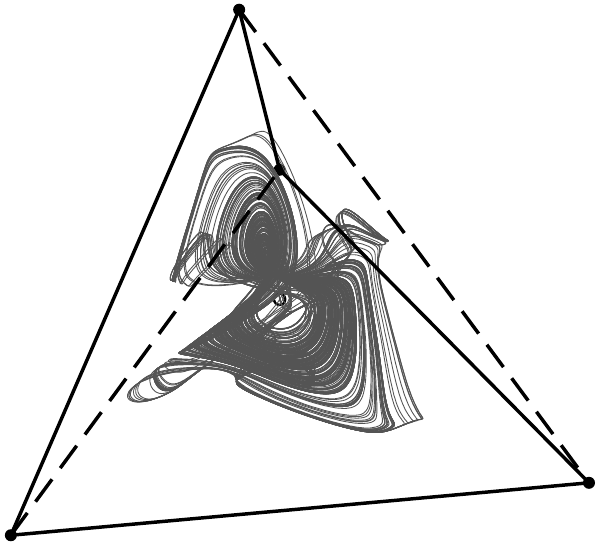}}\quad
\subfigure[\ Periodic orbit near heteroclinic network]{\includegraphics[scale=0.65]{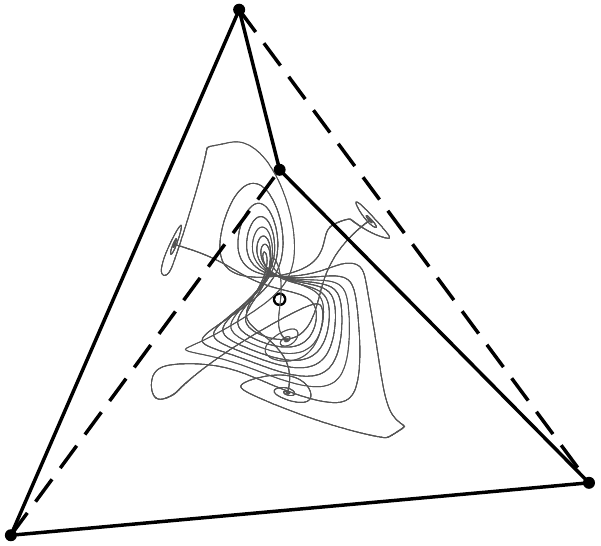}}
\caption{\label{fig:ChaosHetnetN4}Heteroclinic networks organize chaotic behavior in~$\CIR$ for networks of $N=4$ oscillators; line styles on the boundary of~$\CIR$ are as in Figure~\ref{fig:CIR}. The right panel shows a trajectory with positive maximal Lyapunov exponents for phase shift parameters $\chi = (0.154, 0.318, 0, 1.74, 0)$ that comes close to the boundary of~$\CIR$. For nearby parameter values $\chi = (0.2, 0.316, 0, 1.73, 0)$ there is an attracting periodic orbit close to a heteroclinic network involving two saddle equilibria, one a saddle-focus, on the boundary of~$\CIR$.
}
\end{figure}

Trajectories with positive maximal Lyapunov exponents~$\lmax$ also appear close to the boundary of~$\CIR$. Figure~\ref{fig:ChaosHetnetN4}(a) shows a solution $\theta$ in~$\CIR$ for $\chi = (0.154, 0.318, 0, 1.74, 0)$. These solutions appear to be organized by heteroclinic networks that involve saddle-focus equilibria on the boundary of~$\CIR$: Figure~\ref{fig:ChaosHetnetN4}(b) shows a stable periodic orbit close to such a heteroclinic network for parameters $\chi = (0.2, 0.316, 0, 1.73, 0)$. This suggests that chaos can also arise through a Shilnikov saddle-focus scenario~\cite{Shilnikov1965} on the boundary~of~$\CIR$.

\subsection{Chaos in networks of $N=5$ oscillators}

\begin{figure*}
\subfigure[\ Chaotic Fluctuations of~$R$]{\raisebox{0.15cm}{\includegraphics[scale=\imagescaling]{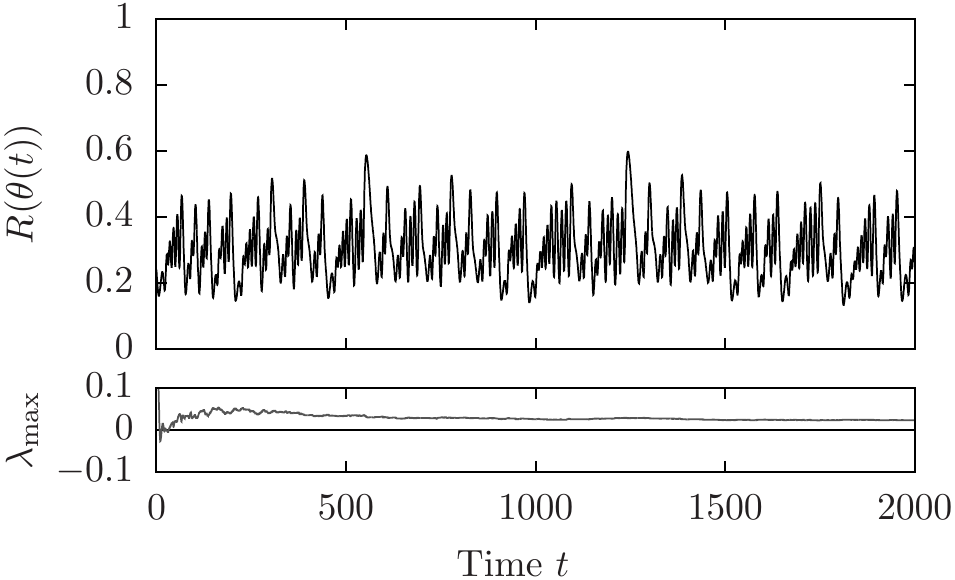}}}\qquad
\subfigure[\ Positive $\lmax$ for $N=5$]{\includegraphics[scale=\imagescaling]{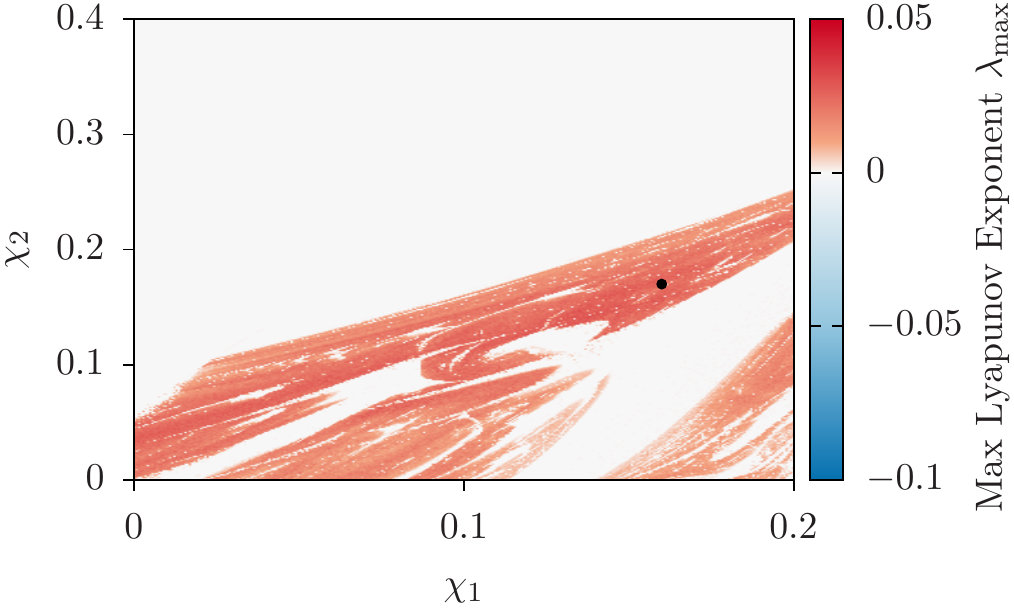}}
\caption{\label{fig:ChaosN5}
Chaotic attractors exist in~$\CIR$ for networks~\eqref{eq:principal} of $N=5$ oscillators. The left panel shows chaotic fluctuations of the order parameter $R(\theta(t))$ and the convergence of the maximal Lyapunov exponent~$\lmax$ for a trajectory for phase shift parameters $\chi = (0.16, 0.17, 0, 1, 0)$. The right panel shows trajectories with chaotic dynamics for varying parameters $\chi= (\chi_1, \chi_2, 0, 1, 0)$.  A black dot indicates the choice of parameters in Panel~(a).}
\end{figure*}

Positive maximal Lyapunov exponents also for networks of $N=5$ oscillators. Figure~\ref{fig:ChaosN5} shows positive maximal Lyapunov exponents and chaotic order parameter fluctuations for the dynamics of~\eqref{eq:principal} for varying phase shift parameters~$\chi$ and fixed initial condition $\theta(0) = (0.646, 1.726, 3.269, 5.295, 2\pi)$. Note that for the same parameter range as in Figure~\ref{fig:ChaosN5}, positive maximal Lyapunov exponents also arise for the dynamics of $N=4$ oscillators (not shown).

\section{Discussion}

We show that symmetrical phase oscillator networks with coupling that involves nonpairwise interaction can exhibit chaotic dynamics with coupling functions that only contain two nontrivial harmonics. As demonstrated in Section~\ref{sec:Chaos}, this is also the case for a network of $N=4$ oscillators, the smallest networks that can support chaotic dynamics. By contrast, for networks of four oscillators with pairwise interaction the only known example of a coupling function that gives rise to chaotic dynamics has four nontrivial harmonics~\cite{Bick2011}. Coupling functions with two harmonics are sufficient for larger networks~\cite{AOWT07}. The emergence of positive Lyapunov exponents for nonpairwise coupling begs to be explored further. Our results suggest that chaos can arise through period doubling and---modulo the symmetry on the invariant region---in a Shilnikov scenario involving a saddle focus on the boundary of the canonical invariant region. However, the boundaries of parameter values giving rise to positive~$\lmax$ apparent in Figures~\ref{fig:ChaosN4}(b) and~\ref{fig:ChaosN5}(b) remain to be traced out. Moreover, positive Lyapunov exponents arise in the same region of parameter space. Is it possible to find a (set of) coupling functions that give chaotic dynamics for any~$N\geq 4$?

Our results clarify the role amplitude dynamics in the emergence of chaos for oscillators beyond the weak coupling limit. While it has been argued that amplitude degrees of freedom are crucial for the emergence of chaotic dynamics in fully symmetric coupled oscillators~\cite{HR92,NK94}, our results suggest that these additional degrees of freedom are not necessary for four or more oscillators: chaos can arise in the phase reduction of symmetrically coupled oscillators close to a Hopf bifurcation in the weak coupling limit through nonpairwise coupling. Note that higher order expansions of the phase dynamics for symmetric oscillators close to a Hopf bifurcation may induce interaction terms involving five or more phases. These may affect the qualitative dynamics of the phase reduction for $N\geq 5$ oscillators but reduce to interactions of four phases in networks of $N=4$ oscillators; cf.~Section~\ref{sec:SmallNets}. By contrast, the invariant torus in the weak coupling limit for three oscillators does not support any chaotic dynamics due to the continuous phase shift symmetry. Thus, chaotic dynamics for three symmetrically coupled oscillators~\cite{AP03} can only occur in the full system for time scales where the weak coupling approximation breaks down.

Nonpairwise interaction between phase oscillators also facilitates the emergence of chaotic weak chimeras, i.e., dynamically invariant sets on which oscillators are locally frequency synchronized. While recent results on the existence of chaotic weak chimeras relied on pairwise interactions and coupling functions with four nontrivial harmonics~\cite{Bick2016}, nonpairwise interaction yields another mechanism to construct such solutions. Moreover, as nonpairwise coupling arises in a phase reduction of more general oscillators our results provide a link between chaotic weak chimeras and chimera states found for more general oscillators beyond the weak coupling limit~\cite{SethiSen14}---see also~\cite{Bick2015}.

Higher order interactions that involve nonpairwise terms also arise in oscillators with mean field coupling~\cite{Komarov2015} and are of interest for applications. In fact, methods for the analysis of time series of oscillatory data explicitly address the problem of reconstructing higher order terms~\cite{Kralemann2011, Stankovski2015}. Thus, we anticipate that a detailed understanding of the dynamical effects induced by general nonpairwise coupling will give additional insights into the analysis of real-world networks of oscillatory units with generic coupling.

\subsection*{Acknowledgements}
The authors would like to thank P.~Glendinning for helpful conversations and M.~Timme, Network Dynamics, and the Department of Nonlinear Dynamics at the Max Planck Institute for Dynamics and Self-Organization in G\"ottingen, Germany for generous access to their high performance computing facilities to perform large scale computations.
CB gratefully acknowledges financial support from the People Programme (Marie Curie Actions) of the European Union's Seventh Framework Programme (FP7/2007--2013) under REA grant agreement no.~626111. 

\bibliographystyle{plain}


\end{document}